\documentstyle[preprint,tighten,aps,prb]{revtex}

\begin{document}
\draft
\title{Exciton and confinement potential effects on the resonant Raman 
scattering in quantum dots}
\author{E. Men\'endez-Proupin\cite{also} and J. L. Pe\~na\cite{also2}}
\address{Centro de Investigaci\'on y Estudios Avanzados del Instituto\\
Polit\'ecnico Nacional, Unidad M\'erida, A.P. 73 ``Cordemex'',\\
C.P. 97310, M\'erida, Yucat\'an, M\'exico}
\author{C. Trallero-Giner}
\address{Universidad de La Habana, Dept. de F\'\i sica Te\'orica\\
Vedado 10400, La Habana, Cuba}
\maketitle

\begin{abstract}
Resonant Raman scattering in spherical semiconductor quantum dots 
 is theoretically investigated.  The
Fr\"{o}hlich-like interaction between electronic states and optical
vibrations has been considered. The Raman profiles are studied for the
following intermediate electronic state models: (I) uncorrelated
electron-hole pairs in the strongly size-dependent quantized regime; 
(II) Wannier-Mott
excitons in an infinite potential well;  (III) excitons in a finite
confinement barrier. It is shown that the finite confinement
barrier height and the
electron-hole correlation determine the absolutes values of the scattering
intensities and substantially modify the Raman line shape, even in the
strong confinement regime.
\end{abstract}

\pacs{63.20.Dj; 68.65.+9; 63.20.-e}


\newpage
\narrowtext

\section{Introduction}

During the last decade, technological developments have made possible the
fabrication of one and zero dimensional nanostructures such as quantum wires
and quantum dots (QDs). The interest on these systems comes from their novel
optical and transport properties and has been stimulated by the success of
quantum wells in technology. The effect of reduced dimensionality on the
electronic excitations and the related optical properties has been the subject of
intensive investigation and nowadays it is more or less well
understood. 

Semiconductor-doped glasses (SDGs) are particularly useful to investigate the
vibrational modes in quasi-zero-dimensional systems, because the use of
appropriate thermal annealing techniques makes it possible to grow
semiconductor nanocrystallites with  small enough radius to show the effects
of spatial confinement on the optical vibrational modes. Raman spectroscopy
is a valuable tool to probe the active optical modes and also to obtain
information about the electronic system. In addition, resonant Raman
scattering (RRS) can be used as a size selective technique\cite{3a}, which
could play an important role on SDGs due to their broad dispersion in
microcrystallite sizes. Recently, the mechanism and features of Raman
scattering by semiconductor nanocrystallites have been studied\cite
{r1,r2,r3,r4}, showing the effects of the reduced dimensionality on the
Raman shift and lineshape. A preliminary theory of first-order RRS in
spherical microcrystallites has been developed in \onlinecite{r1}
and \onlinecite{chamb} on the basis of a continuum model for polar optical
vibrations. These models consider the electronic intermediate states as 
uncorrelated electron-hole pair (EHP) states, that is, in the strong size
quantized regime (model I).
An extension to the above theories, considering the
electron-hole interaction effects has been recently
presented\cite{eduard} (model II).
The calculations performed in \onlinecite{eduard} are strictly valid
for excitons completely confined within dots. 
Raman scattering in the Fr\"ohlich configuration
considerably depends on the differences between electron and hole
wave functions ({\it electron-hole
decompensation})\cite{lscatter}. The theoretical values 
of the Raman cross section and lineshape should be modified by the
electron-hole model and the confinement potential used in the entire
calculation. Hence, in the framework of a free EHP model with infinite
barriers the same wave functions for electrons and holes are obtained and
null contribution of the Fr\"ohlich mechanism to the Raman cross section is
achieved. The scattering efficiencies following models I and II
considerably differ when absolute values are calculated, even for
QDs with radii smaller than the exciton Bohr radius. In
model I the finite confinement barriers are considered but
regardless of excitonic effects. Model II includes the
electron-hole correlation in an infinite barrier, but the chosen potential
diminishes the electron-hole decompensation occurring through the
finite band offsets potential. In \onlinecite{eduard} on
the lines of model II an effective radius $R_{ef}$ was introduced
in order to take into account the penetration of the exciton
wave function into the adjacent medium. This procedure allows, in some way,
the RRS calculations in real systems using the mathematical simplicity of
the infinite barrier basis functions. 
We will show that within the above approach accurate exciton
ground state energies can be achieved. This approach  underestimates the 
calculated Raman absolute values. It is well established 
that a reliable Raman scattering theory becomes necessary in order to
interpret RRS absolute values in semiconductors\cite{cant}.
The purpose of the present paper is to clarify 
the electron-hole decompensation effect
on the absolute values of scattering intensities and Raman
lineshapes taking into account uncorrelated and correlated
electron-hole theories and using different confinement potential models.

The paper is organized as follows. In Sec.~II we provide the
theoretical basis needed to obtain the Raman cross section where
the electronic intermediate states are excitons in a finite
spherical potential box (model III). Theories I and II are
derived as proper limits from the more general model III. We
also compare the Raman intensity values for  CdS QDs embedded
in a glass matrix, obtained along the lines of the above described
theoretical models. In Sec.~III we present the conclusions of the
present work.

\section{Results and discussion}

The Raman cross section $\partial^2 \sigma/\partial\Omega%
\partial\omega_s$ of a dot of radius $R$ can be expressed as\cite{eduard}
\begin{eqnarray}  \label{ec:2}
\frac{\partial^2 \sigma}{\partial\Omega\partial\omega_s}=S_0\sum_{n_p}
\left|\sum_{N,N^{\prime}}\frac{f_N \langle
N|h_{E-P}^{(n_p)}|N^{\prime}\rangle f_{N^{\prime}}} {(\hbar\omega_s-E_{N^{%
\prime}}(R)+i\Gamma_{N^{\prime}})(\hbar\omega_l-E_{N}(R)+i\Gamma_{N})}
\right |^2 \times  \nonumber \\
\frac{\Gamma_{n_p}/\pi}{(\hbar\omega_l-\hbar\omega_s-\hbar%
\omega_{n_p}(R))^2+\Gamma_{n_p}^2}.
\end{eqnarray}
Here, $\hbar\omega_l$ ($\hbar\omega_s$) is the incoming (outgoing)
 photon energy, $E_N$ ($\Gamma_N$) is the energy (broadening) of the
intermediate $L=0$ electronic state $|N\rangle$ ($L$ being the quantum
number of the total electronic angular momentum square), $f_N$ their optical
strengths, $\langle N|h_{E-P}^{(n_p)}|N^{\prime}\rangle$ is the matrix element
of the electron-Fr\"{o}hlich-type lattice interaction (in dimensionless units%
\cite{eduard}) and $n_p$ is the vibron\cite{t13}  quantum number with
angular momentum $l_p=0$ and frequency $\omega_{n_p}$. $S_0$ is a constant
which depends on the semiconductor parameters and the embedding medium%
\cite{eduard}.

The exciton wave function $\Psi({\bf r}_e,{\bf r}_h)$ is obtained by the
expansion 
\begin{equation}
\Psi _{N,L,M}({\bf r}_e,{\bf r}_h)=\sum_{\alpha =\{n_e,n_h,l_e,l_h\}}C_{N,L,M}(\alpha
)\Phi _\alpha ({\bf r}_e,{\bf r}_h),  \label{exc:funcion}
\end{equation}
where the basis functions $\Phi_\alpha ({\bf {r}_e,{r}_h)}$ are
eigenfunctions of the total angular momentum square $\hat L^2$, its z-projection $%
\hat L_z$, and the Hamiltonian of the free EHP in the dot. The functions $%
\Phi _\alpha ({\bf r}_e,{\bf r}_h)$ are constructed from the dot electron
and hole wave functions ($\phi _{n_e,l_e,m_e}({\bf r}_e)$ and $\phi
_{n_h,l_h,m_h}({\bf r}_h)$), through the relation 
\begin{equation}
\Phi _\alpha ({\bf r}_e,{\bf r}_h)=\sum_{m_e,m_h}(l_el_hm_em_h|LM)\phi
_{n_e,l_e,m_e}({\bf r}_e)\phi _{n_h,l_h,m_h}({\bf r}_h).  \label{ocho}
\end{equation}
$(l_el_hm_em_h|LM)$ being the well known Clebsch-Gordon coefficients. 

The coefficients $C_{N,L,M}(\alpha )$ and the eigenenergy $E_N$ are obtained from
numerical diagonalization of the exciton Hamiltonian in a spherical potential well, 
using the basis defined by 
equation~(\ref{ocho})\cite{nota1}. If the uncorrelated theory
(model I) is considered, for every eigenstate there is only one 
 non-zero coefficient  $C_{N,L,M}(\alpha )$ in 
the expansion (\ref{exc:funcion}). This approach leads to the same results
as the formalism of ~\onlinecite{chamb}.
On the other hand, models II and III differ upon the
 radial parts of the electronic wave functions
$\phi _{n_e,l_e,m_e}({\bf r}_e)$ and $\phi_{n_h,l_h,m_h}({\bf r}_h)$, which 
depend on the chosen confinement potential.

The resonance condition with a particular electronic level $N$ is given by
the equations $\hbar \omega _s=E_N(R)$ (outgoing resonance) or $\hbar \omega
_l=E_N(R)$ (incoming resonance). In the dipole approximation only excitons
with $L=0$ are created or annihilated, corresponding to $l_e=l_h$ interband
transitions in the free EHP model. If the valence band mixing is neglected,
only $l_p=0$ vibrons contribute to the Raman scattering. 

The calculation of the matrix elements of Eq.~(\ref{ec:2}) has been
performed in \onlinecite{eduard} for the case of totally confined
excitons, while the strong size quantized regime (non exciton effects) has 
been developed in \onlinecite{chamb}. The parameters used in
our calculations  correspond to a CdS QD of
radius $20$~\AA{}\cite{eduard}. This means that the dot is in the strong confinement
regime. In this regime the Coulomb attraction shifts the EHP energies to
lower values and small changes on the wave functions are expected. 

Figure~\ref{fig1} shows the electron and hole density of probability for the
three lower $L=0$ excitonic eigenstates as functions of  $r$,
the distance to the 
dot center. The density of probability in the case of the uncorrelated
EHP model (I) is  shown by dashed curve. For the $N=1,L=0$ excitonic state
the effect of correlation is to push both the electron and the hole to the
dot center. As can be seen, for the system under consideration 
(CdS QD of radius $20$~\AA ) the effect of
the finite confinement on the electron-hole decompensation is larger than
that of the electron-hole interaction. As we shall see, if 
the former effect is neglected considerable changes 
on the predicted Raman cross section absolute values are obtained. 
In Figure~\ref{fig2}(a) we compare the calculated Raman cross-
section for incoming light in resonance with the $N=1$ excitonic
state following models I, II and III. The incoming resonances 
happen at $\hbar\omega _l=2.870$ eV in the finite barrier
excitonic model III 
(solid curve), at $\hbar\omega _l=3.014$ eV in the 
uncorrelated EHP model I 
(dashed curve) and at $\hbar \omega _l=2.878$ eV for the excitonic model II 
(dot-dashed curve), assuming an effective radius
$R_{ef}=26$~\AA{} to simulate the finite-barrier height. It can be seen that
accurate $N=1$ exciton energy can be obtained following the formalism of
model II. 
 The $N=1$ excitonic state, as can be seen in Table~\ref{tab2}(a),
is mainly composed of EHP states with quantum numbers $n_e=n_h=1,l_e=l_h=0$
with a large oscillator strength $|f_N|^2$, giving the main contribution to
the cross-section in the resonance condition. The line shape is almost the same
in the three models. The difference between those 
 models lays in the absolute values of the cross-section, which is 
smaller in model II. It is clear that the dominant effect on absolute
values comes from: (a) the values of the oscillator strength\cite{11};
(b) the EHP wave functions decompensation produced by the finite 
depth of the spherical well. It can be seen from Figure~\ref{fig1}
that the electron-hole decompensation for the first level is
slightly larger in model III than the free EHP theory (I), something that is
reflected on the values of the exciton-vibron matrix elements reported in
Table~\ref{tab2}. In
the case of excitons completely confined 
(II) the exciton-vibron matrix elements 
$\langle 1|h^{(n_p)}|1\rangle$ are one order of magnitude
smaller than I and III (see \onlinecite{eduard}). 
However, we must note that in the case of the
electrons, the effective mass in the glass matrix is five times larger than
its value inside the dot, causing
an extremely large 
decompensation. A similarly large effect can be achieved if one of the
barrier heights is too small.

Figure~\ref{fig2}(b) shows the Raman spectrum in the case of incoming
resonance with the $N=2$ exciton at 
$\hbar\omega_l=3.439$, $3.205$ and $3.292$~eV in models I, II and
III respectively. In I, the $N=2$ state is 
the free EHP with quantum numbers $n_e=1,n_h=2,l_e=l_h=0$ and it has a weak
optical activity, as can be seen from the corresponding oscillator strength $%
|f_N|^2$ in Table~\ref{tab2}(b). Nevertheless, the excitonic effects produced
by the Coulomb interaction greatly enhance its oscillator strength (see
Table~\ref{tab2}(a)) and a strong incoming resonance is obtained. It must be
noted that even when the matrix element $\langle 2|h^{(n_p)}|2\rangle$ is
maximum for $n_p=2$, the main contribution to the cross-section in Figure~\ref
{fig2}(b) corresponds to $n_p=1$, a fact that can be explained by
interference effects due to virtual transitions between $N=2$ and $N=3$
excitonic levels. Figure~\ref{fig1}(b) shows that electron-hole
decompensation is similar for I and III. Nevertheless a
completely confined exciton theory with an effective radius
gives a matrix elements $\langle 2|h^{(n_p)}|2\rangle$ one order
of magnitude smaller than those reported in Table~\ref{tab2}.

Figure~\ref{fig2}(c) shows the spectrum for the case of incoming resonance
with the $N=3$ level, at $\hbar\omega_l=3.479$, $3.310$ and $3.339$ eV in
models I, II and III, respectively. The results of the theories considered here
 present great differences: (a) Model II predicts a cross-
section smaller than that for the $N=1$ incoming resonance (Fig.~\ref{fig2}%
(a)), while I and III predict larger cross-sections  than those of
Fig.~\ref{fig2}(a). (b) In model III, the peak associated to the $n_p=2$
vibron becomes bigger  than the $n_p=1$ peak. In models I and III, because 
the energy of incoming resonance $\hbar\omega_l=E_3$ is very close to the
energy of outgoing resonance with the $N=2$ excitonic state $%
\hbar\omega_s\simeq E_2+\hbar\omega_p$ (see Table~\ref{tab2}), a 
quasi-double resonant condition takes 
place in the scattering process. Hence, the
Raman cross-section values are strongly dependent on the matrix elements $%
\langle 3|h^{(n_p)}|2\rangle$, which are maximum for $n_p=2$, explaining why
the $n_p=2$ peak is greatly enhanced. Moreover, the $n_p=1$ contribution is
dropped because of interference effects between $N=2$ and $N=3$ excitonic
transitions mediated by the matrix elements $\langle 3|h^{(n_p)} |3\rangle$
and $\langle 3|h^{(n_p)}|2\rangle$. Owing to symmetry, 
the matrix element $\langle 3|h^{(n_p)}|2\rangle$ vanishes in the framework
of the free EHP model  and the quasi-double resonance effect is not observed
in Figure~\ref{fig2}(c). We have also calculated the spectrum
 in the outgoing resonance with $N=3$, using models I and
III. In this case the double-resonance condition is not fulfilled and the
obtained cross-section is similar in both models.

We have finally compared the integral Raman intensity for the $n_p=1$
vibron of a 20~\AA{} CdS QD and it is shown in Figure~\ref{fig3} as a
function of the incident photon energy. We have used the same broadening of $%
\Gamma=5$ meV for all the excitonic levels. This plot takes up the effects
already presented in previous figures over the absolute values of the Raman
 spectra. The red shift of the resonances due to the attractive
electron hole interaction is shown. Due to the small optical oscillator
strength the intensities corresponding to the incoming and outgoing
resonances with the second EHP level in model I, are insignificant compared
to those of the first and third levels. Model III predicts stronger resonances
for the $N=1$ exciton than model I, a fact explained by the enhancement of
its oscillator strength. For all models the $N=1$ outgoing resonance is
stronger than the incoming one, but for the $N=2$ state the opposite is
obtained. The above feature is a general result of the Fr\"{o}hlich-like
interaction in a quantum dot. The $N=2$ excitonic state has an oscillator
strength equal to 1.08, a factor about 30 times larger than for the free EHP
(see Table~\ref{tab2}) and this is the cause of the strong $N=2$ incoming
resonance seen in the plot. The outgoing peak for the $N=3$ level is smaller
in model III than that of the free EHP theory. This is explained by the reduction
of the electron-hole decompensation observed in model III (see
Figure~\ref{fig1}(c)).
The intensities calculated according to model
II are two orders of magnitude smaller than that of
the exciton in the finite-barrier models. This calculation is not presented
in Fig.~\ref{fig3}.

\section{Conclusions}

We have studied the influence of excitonic and finite confinement effects
on the first-order Raman cross-sections for longitudinal
optical vibrons in nanospherical semiconductor quantum dots.
 We have compared the
 predictions of three models for the intermediate electronic states: 
(I) uncorrelated electron hole pairs with finite dot confinement; 
(II) excitons completely confined in an spherical box with an effective
radius; (III)
excitons in a finite confinement barrier. 
The main conclusion of the present
work is that the Raman spectra and the resonance profile 
absolute values for the Fr\"{o}hlich-type-interaction Hamiltonian 
in QDs should be predicted by a theory that takes into consideration both 
the finite confinement barrier height and electron-hole
correlation effects. Even in the strong quantum confinement regime excitons
and the conduction and valence-band offsets substantially modify the
features of the resonant Raman spectra, particularly in presence of
quasi-double resonances.

\acknowledgements

We would like to thank F. Comas and J. Tutor for critical readings of the manuscript.
Two of us (E. M.-P. and J. L. P.) would like to thank the
support of the Secretary of Public Education of Yucatan State, Mexico.

\newpage
\begin{figure}[tbp]
\caption{ Density of probability function for the hole (curve A) $4\pi
r^2\int |\Psi_{N}({\bf r}_e,{\bf r})|^2d^3{\bf r}_e$ and for the electron
(curve B) $4\pi r^2\int |\Psi_{N}({\bf r},{\bf r}_h)|^2d^3{\bf r}_h$ , for
the states $N=1,2,3$ and $L=0$ considering finite band offsets. The solid
curves correspond to calculations using the excitonic model and the dashed
curves, using the free electron-hole pair model. }
\label{fig1}
\end{figure}
\begin{figure}[tbp]
\caption{ Raman cross-section of a 20~\AA{} CdS quantum dot, calculated using
different electronic models: 
(I) uncorrelated EHP intermediate states (dashed curve);
(II) excitonic intermediate states in a spherical box with an effective radius
$R_{ef}=26$~\AA{} (dot-dashed curve);  
(III) excitonic intermediate states (solid curve). 
a) At $\hbar\omega_l=3.014$, $2.878$ and $2.870$ eV for model 
I, II and III respectively (incoming resonance with $N=1$, $L=0$ EHP); 
b) at $\hbar\omega_l=3.439$, $3.205$ and $3.292$ eV for model
I, II and III respectively (incoming resonance with $N=2$,
$L=0$ EHP); 
c) at $\hbar\omega_l=3.479$, $3.310$ and $3.339$ eV
for model I, II and III respectively (incoming resonance with $N=3$, $L=0$
EHP).}
\label{fig2}
\end{figure}
\begin{figure}[tbp]
\caption{ Raman intensity for a 20~\AA{} CdS quantum dot, as a function of the
incident photon energy, according to model III (solid curve) and I (dashed
curve). }
\label{fig3}
\end{figure}
\newpage
\widetext
\begin{table}[tbp]
\caption{Values of the coefficients $C_{N,0,0}$, resonance energies $E_N$
(incoming) and $E_N+\hbar\omega_p$ (outgoing), oscillator strength $|f_N|^2$
and dimensionless exciton-lattice matrix elements $\langle N
|h_{E-P}^{(n_p)}|N'\rangle$ for different $n_p$ vibronic modes contributing
to the Raman cross-section, calculated using the excitonic and the free
electron-hole pair models with finite barriers and the parameters of
~\protect\onlinecite{eduard}.
}
\label{tab2}
a) Excitonic model 
\begin{tabular}{||ccccccc||}
$N$ & $C_{N,0,0}(n_e,n_h,l_e,l_h)^2$ & $E_N(eV)$ & $|f_N|^2$ & $n_p$ & $\langle
N|h_{E-P}^{(n_p)}|N\rangle $ & $\langle N |h_{E-P}^{(n_p)}|N+1\rangle$ \\ 
&  & ($E_N+\hbar\omega_p$) &  &  &  &  \\ \hline
\  &  & 2.870 &  & 1 & $-9.5\times 10^{-2}$ & $2.3\times 10^{-1}$ \\ 
1 & $C(1,1,0,0)^2=0.98$ & (2.908) & 1.67 & 2 & $-8.4\times 10^{-3}$ & $%
9.4\times 10^{-2}$ \\ 
&  &  &  & 3 & $-8.1\times 10^{-5}$ & $-1.3\times10^{-3}$ \\ \hline
& $C(1,2,0,0)^2=0.83$ & 3.292 &  & 1 & $-1.1\times 10^{-1}$ & $3.8\times
10^{-3}$ \\ 
2 & $C(1,1,1,1)^2=0.16$ & (3.329) & 1.08 & 2 & $-1.4\times 10^{-1}$ & $%
-6.2\times 10^{-2}$ \\ 
&  &  &  & 3 & $-5.0\times 10^{-2}$ & $-1.8\times 10^{-2}$ \\ \hline
& $C(1,2,0,0)^2=0.16$ & 3.339 &  & 1 & $-5.9\times 10^{-2}$ & $-1.1\times
10^{-1}$ \\ 
3 & $C(1,1,1,1)^2=0.82$ & (3.376) & 2.74 & 2 & $4.0\times 10^{-3}$ & $%
4.2\times 10^{-3}$ \\ 
&  &  &  & 3 & $8.3\times 10^{-3}$ & $2.1\times 10^{-2}$%
\end{tabular}
\end{table}

\begin{table}[tbp]
b) Free electron hole pair model 
\begin{tabular}{||ccccccc||}
$N$ & $C_{N,0,0}(n_e,n_h,l_e,l_h)^2$ & $E_N(eV)$ & $|f_N|^2$ & $n_p$ & $\langle
N|h_{E-P}^{(n_p)}|N\rangle $ & $\langle N |h_{E-P}^{(n_p)}|N+1\rangle$ \\ 
&  & ($E_N+\hbar\omega_p$) &  &  &  &  \\ \hline
&  & 3.014 &  & 1 & $-8.6\times10^{-2}$ & $-2.6\times10^{-1}$ \\ 
1 & $C(1,1,0,0)^2=1$ & (3.051) & 0.96 & 2 & $-3.0\times10^{-3}$ & $%
-9.2\times10^{-2}$ \\ 
&  &  &  & 3 & $-2.8\times 10^{-4}$ & $-5.3\times 10^{-3}$ \\ \hline
&  & 3.439 &  & 1 & $-1.2\times 10^{-1}$ & 0 \\ 
2 & $C(1,2,0,0)^2=1$ & (3.476) & 0.03 & 2 & $-1.7\times 10^{-1}$ & 0 \\ 
&  &  &  & 3 & $-5.0\times 10^{-2}$ & 0 \\ \hline
&  & 3.479 &  & 1 & $-7.0\times 10^{-2}$ & 0 \\ 
3 & $C(1,1,1,1)^2=1$ & (3.516) & 2.87 & 2 & $2.6\times 10^{-2}$ & 0 \\ 
&  &  &  & 3 & $-1.9\times 10^{-3}$ & 0 \\ 
&  &  &  &  &  & 
\end{tabular}
\end{table}


\begin{references}
\bibitem[*]{also}  Also at Universidad de La Habana, Dept. de F\'\i sica
Te\'orica, Vedado 10400, La Habana, Cuba

\bibitem[\dag]{also2} Also at Centro de Investigaci\'on en 
Ciencia Aplicada y Tecnolog\'{\i}a Avanzada-IPN, Legaria 694, 
C.P. 11500, Mexico, D.F. 

\bibitem{3a}  C. Trallero-Giner, A. Debernardi, M. Cardona, and E.
Men\'endez-Proupin, Phys. Rev. B {\bf 57}, 4664 (1998).

\bibitem{r1}  M. C. Klein, F. Hache, D. Ricard, and C. Flytzanis, Phys. Rev.
B {\bf 42}, 11123 (1990).

\bibitem{r2}  G. Scamarcio, M. Lugara, and D. Manno, Phys. Rev. B {\bf 45},
13792 (1992).

\bibitem{r3}  W. S. O. Rodden, C. M. Sotomayor-Torres, and C. N. Ironside,
Semicond. Sci. Technol. {\bf 10}, 807 (1995).

\bibitem{r4}  T. D. Krauss, F. W. Wise, and D. B. Tanner, Phys. Rev. Lett. 
{\bf 76}, 1376 (1996).

\bibitem{chamb}  M. P. Chamberlain, C. Trallero-Giner, and M. Cardona, Phys.
Rev. B {\bf 51}, 1680 (1995).

\bibitem{eduard}  E. Menendez, C. Trallero-Giner, and M. Cardona,  Phys.
Status Solidi (b) {\bf 199}, 81 (1997); {\it ibid}. {\bf 201}, 551 (1997).

\bibitem{lscatter}  M. Cardona, in {\it Light Scattering in Solids II}, ed.
by M. Cardona and G. G\"untherodt, Topics in Applied Physics, Vol. 50
(Springer, Heidelberg, 1982), p. 19.

\bibitem{cant} C. Trallero-Giner, A. Cantarero, and M. Cardona, Phys. Rev.
B {\bf 40}, 4030 (1989)

\bibitem{t13}  In a quantum dot is lost the concept of phonon as a
quasiparticle with translational symmetry, then it is called vibron.

\bibitem{nota1}  The matrix elements of the exciton Hamiltonian 
in a spherical quantum dot can be found
 in Ref.~\onlinecite{eduard}. A misprint has recently been noted
in Eq. (22) where 
$(-1)^{L+l_h+l_h^{\prime }}$ must appear in place of $(-1)^{L+l_e+l_h}$.

\bibitem{11} The oscillator strength values corresponding to model II are
: $|f_1|^2=1.829$, $|f_2|^2=0.129$, and $|f_3|^2=4.045$. Note the mistaken
values reported in Ref.~\onlinecite{eduard}.

\end{references}
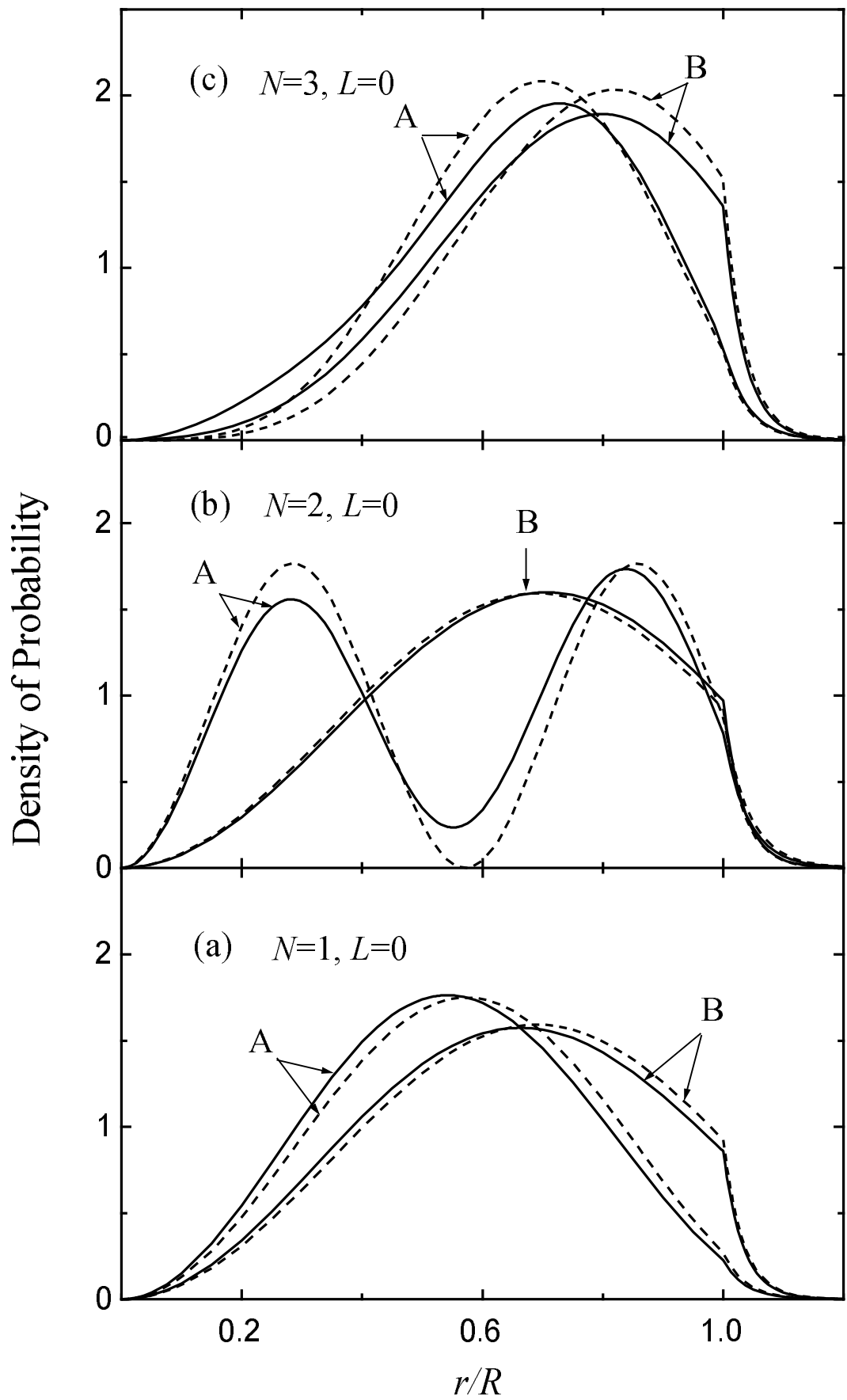
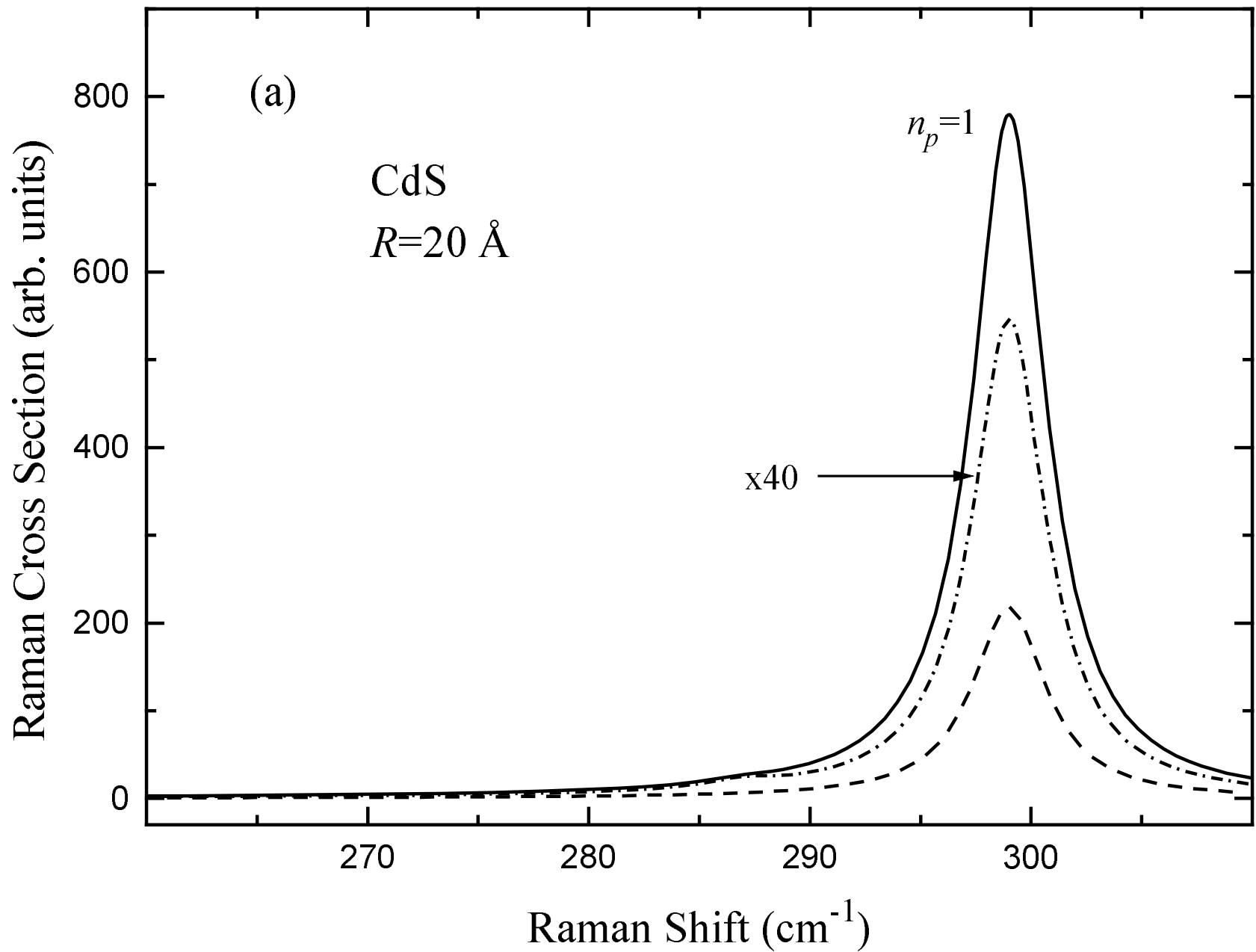
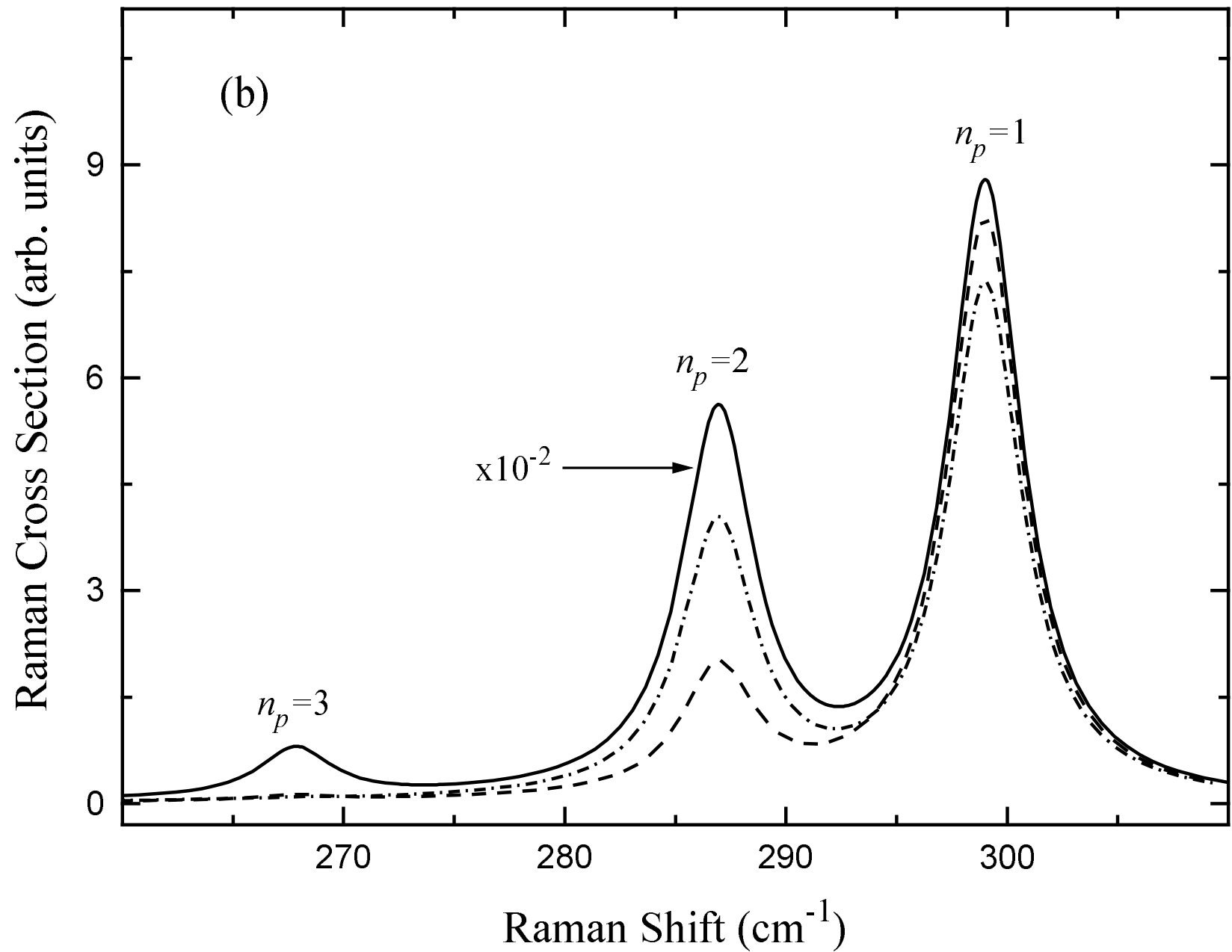
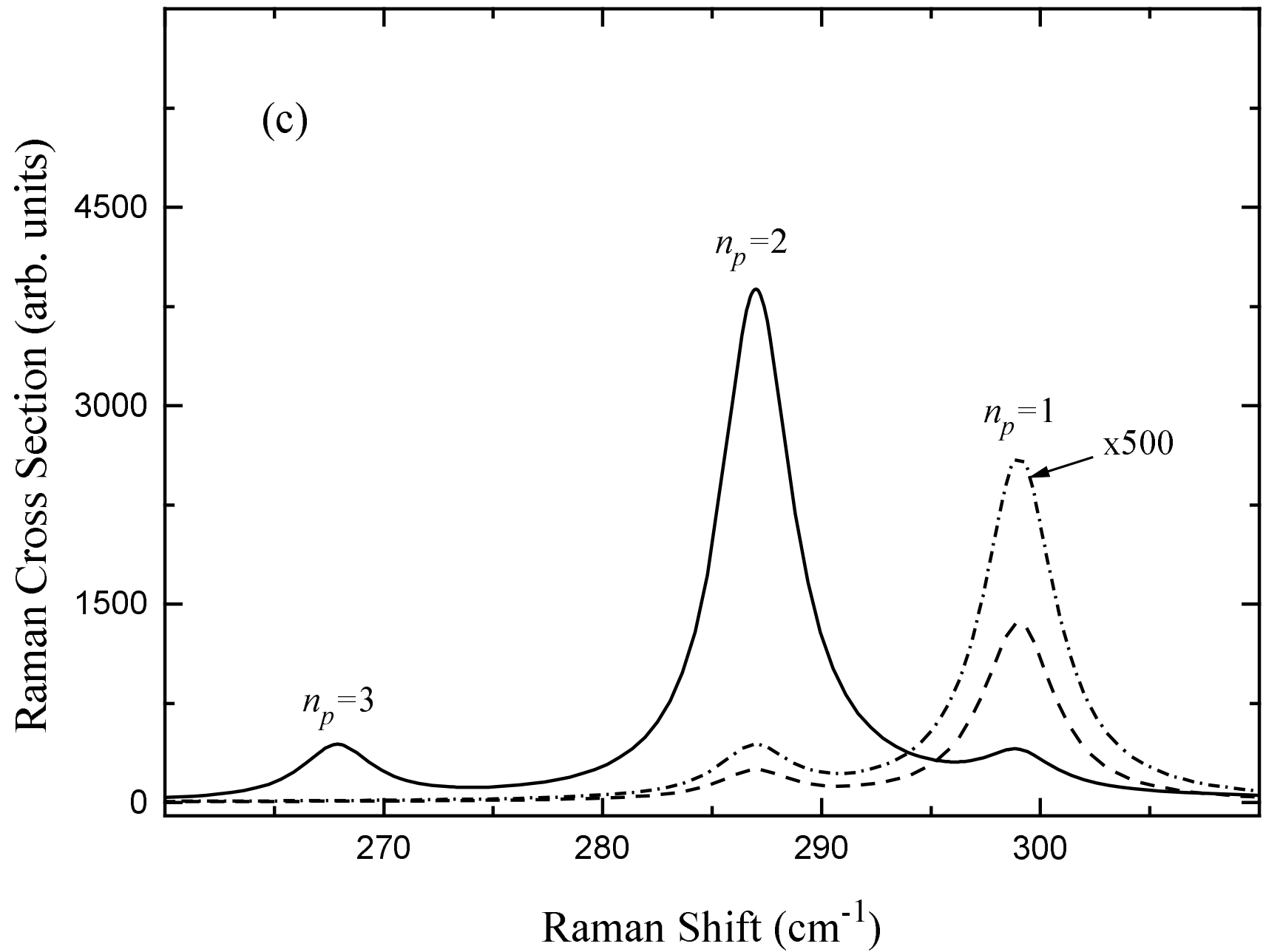
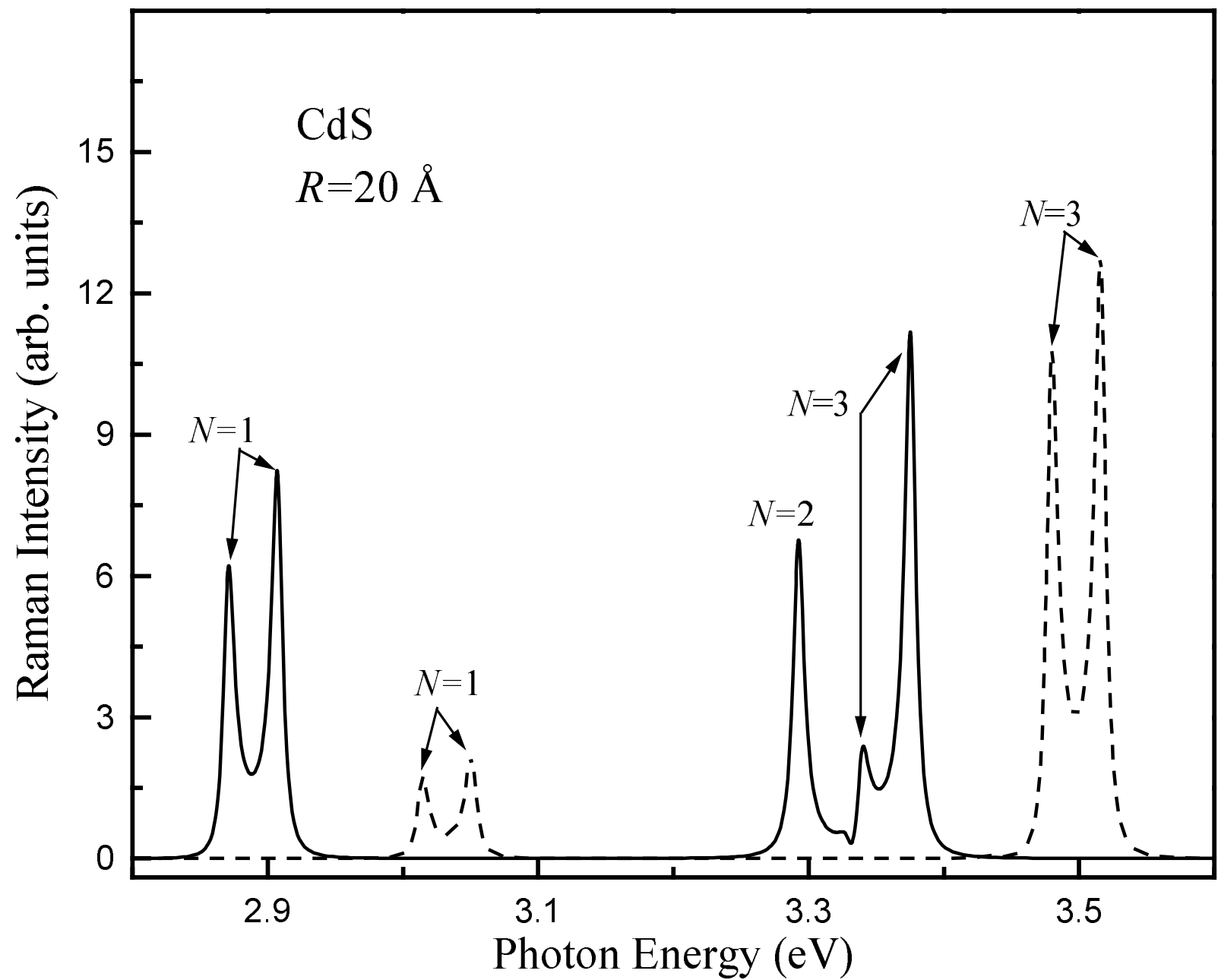
\end{document}